\documentclass[usenatbib,onecolumn]{mn2e}

\def\ergcc{\hbox{\rm\hskip.35em  erg cm}$^{-3}$}

\def\eg{{\it e.g.}}
\def\lap{\hbox{${_{\displaystyle<}\atop^{\displaystyle\sim}}$}}
\def\gap{\hbox{${_{\displaystyle>}\atop^{\displaystyle\sim}}$}}

\newcommand{\ev}[1]{\ensuremath{\langle #1\rangle}}

\newcommand{\ubf}       {\mbox{\boldmath$u$}}

\newcommand{\Bbf}       {\mbox{\boldmath$B$}}

\def\eg{{\it e.g.}}
\def\lap{\hbox{${_{\displaystyle<}\atop^{\displaystyle\sim}}$}}
\def\gap{\hbox{${_{\displaystyle>}\atop^{\displaystyle\sim}}$}}

\usepackage{graphicx}
\usepackage{subfigure}

\title[Torsional oscillations of a magnetar]
{Torsional oscillations of a magnetar with a tangled magnetic field}

\author[B. Link and C. A van Eysden]{Bennett Link\thanks{E-mail:
link@physics.montana.edu} and C. Anthony van Eysden\thanks{E-mail: 
anthonyvaneysden@montana.edu}\\
Department of Physics, Montana State University, Bozeman, Montana, 
59717, USA \\
}

\begin{document}

\date{\today}

\pagerange{\pageref{firstpage}--\pageref{lastpage}} \pubyear{2015}

\maketitle

\label{firstpage}

\begin{abstract}

We propose a scenario for the quasi-periodic oscillations observed in
magnetar flares wherein a tangled component of the stellar magnetic
field introduces nearly isotropic stress that gives the fluid core of
the star an effective shear modulus. In a simple, illustrative model
of constant density, the tangled field eliminates the problematic
Alfv\'en continuum that would exist in the stellar core for an
organized field. For a tangled field energy density comparable to that
inferred from the measured dipole fields of $\sim 10^{15}$ G in SGRs
1806-20 and 1900+14, torsional modes exist with fundamental
frequencies of about 20 Hz, and mode spacings of $\sim 10$ Hz.  For
fixed stellar mass and radius, the model has only one free parameter,
and can account for {\em every} observed QPO under 160 Hz to within 3
Hz for both SGRs 1806-20 and 1900+14. The combined effects of
stratification and crust stresses generally decrease the frequencies
of torsional oscillations by $<10$\% for overtones and increase the
lowest-frequency fundamentals by up to 50\%, and so the star can be
treated as having constant density to a generally good first
approximation. We address the issue of mode excitation by sudden
readjustment of the stellar magnetosphere. While the total energy in
excited modes is well within the energy budget of giant flares, the
surface amplitude is $\lap 10^{-3}$ of the stellar radius for global
oscillations, and decreases strongly with mode frequency. The 626 Hz
QPO reported for SGR 1806-20 is particularly problematic to excite
beyond a surface amplitude of $10^{-6}$ of the stellar radius.

\end{abstract}


\section{Introduction}

Soft-gamma repeaters (SGRs) are strongly-magnetized neutron stars with
magnetic fields of $B=10^{14}-10^{15}$ G that produce frequent,
short-duration bursts ($\lap 1$ s) of $\lap 10^{41}$ ergs in hard
x-ray and soft gamma-rays. SGRs occasionally
produce giant flares that last $\sim 100$ s; the first giant flare to
be detected occurred in SGR 0526-66 on 5 March, 1979
\citep{barat_etal79,mazets_etal79,cline_etal80}, releasing 
$\sim 2\times 10^{45}$ erg \citep{fenimore_etal96}. The August
27th 1998 giant flare from SGR 1900+14 liberated $\gap 4\times
10^{43}$ erg, with a rise time of $<4$ ms
\citep{hurley_etal99,feroci_etal99}. The duration of the initial peak
was $\sim 1$ s \citep{hurley_etal99}. On December 27, 2004, SGR
1806-20 produced the largest flare yet recorded, with a total energy
yield of $\gap 4\times 10^{46}$ ergs.\footnote{These energy estimates
assume isotropic emission.} In both short bursts and in giant flares,
the peak luminosity is reached in under 10 ms.  Measured spin down
parameters imply surface dipole fields of $6\times 10^{14}$ G for SGR
0526-66 \citep{tiengo_etal09}, $7\times 10^{14}$ G for SGR 1900+14
\citep{mereghetti_etal06}, and $2\times 10^{15}$ G for SGR 1806-20
\citep{nakagawa_etal08}, establishing these objects as magnetars.

The giant flares in SGR 1806-20 (hereafter SGR 1806) and SGR 1900+14
(hereafter SGR 1900) showed rotationally phase-dependent,
quasi-periodic oscillations (QPOs). QPOs in SGR 1806 were detected
at 18 Hz, 26 Hz, 30 Hz, 93 Hz, 150 Hz, 626 Hz, and 1837 Hz
\citep{israel_etal05,ws06,sw06,hambaryan_etal11}. 
QPOs in the giant flare of SGR 1900 were detected at 28 Hz, 53 Hz, 84
Hz, and 155 Hz
\citep{sw05}. 
Recently, oscillations at 57 Hz were identified in the
short bursts of SGR 1806 \citep{huppenkothen_etal14a}, and
at 93 Hz, 127 Hz, and possibly 260 Hz in SGR J1550-5418
\citep{huppenkothen_etal14b}.\footnote{\citet{elib10} reported
evidence for oscillations in the
short, recurring bursts of SGR 1806, but this analysis was shown by
\citet{huppenkothen_etal13} to be flawed.}
Overall, the data show that SGRs 1806 and 1900 support oscillations
with spectra that begin at about 20 Hz, with a spacing of some tens of Hz
below 160 Hz. The relative dearth of high-frequency QPOs could
indicate a sparse spectrum above 160 Hz, or highly-preferential
mode excitation. Given that we do not understand how
surface oscillations affect magnetospheric emission, it could be that
some modes are not seen because they do not create large changes in
the x-ray emissivity for our viewing angle.

\citet{duncan98} predicted that the most energetic magnetar
flares would excite observable seismic modes of the neutron star
crust. The easiest modes to excite, Duncan argued, would be torsional
oscillations which, for the assumed composition, begin at
about 30 Hz. The fundamental torsional mode consists of periodic
twisting of two hemispheres in opposite directions. Crust rigidity
provides the restoring force, and the star oscillates as a torsion
pendulum.

Initial theoretical study of the observed QPOs treated the crust
and core as uncoupled for simplicity
\citep{piro05,sa07,lee07,sotani_etal07,wr07,sotani_etal08a,sw09}.
As pointed out by \citet{levin06}, however, the strong magnetic field
of a magnetar introduces an essential complication; the charged
component of the fluid core, primarily protons and electrons, supports
Alfv\'en waves that couple to the highly-conductive crust and affect
the mode spectrum.  A neutron star therefore cannot be regarded as
possessing crust modes that are separate from the core, rather, we
must consider {\em global} oscillations of the crust-core system
\citep{levin06,gsa06}.

It has become clear that the nature of the Alfv\'en wave spectrum of
the core plays a crucial role in the global magneto-elastic
problem. For smooth magnetic field geometries, such as uniform
magnetization or a dipole field, the core fluid can be regarded as
supporting an {\em Alfv\'en continuum} of frequencies; each frequency
corresponds to a natural frequency of a magnetic field line. Since
there is a continuous distribution of line lengths through some range,
the spectrum is continuous. Dipolar field geometries and their
extensions generally give a core spectrum with continuous {\em bands}
separated by {\em gaps}, and a low-frequency cut-off or {\em edge},
much like electronic band structure in solids. The crust, by contrast,
supports only a discrete spectrum of torsional modes if isolated from
the core.  When the crust is included and coupled to the core fluid
through magnetic stresses, the structure of the core spectrum leads to
very rich dynamics as shown by \citet{levin07} for a simple model with
a thin crust and a core of constant magnetization. If energy is
deposited in the crust at one of the natural frequencies of the crust,
and this frequency lies within a portion of the core continuum, the
energy is lost to the core continuum in less than 0.1 s as the entire
core continuum is excited. The crust excitation is effectively damped
through {\em resonant absorption}, a familiar process in MHD; see
\eg, \citet{gp04}.

The turning points and edges in the Alfv\'en continuum play a
special role in the dynamics; power rests at these points and gives
QPOs. Subsequent work on purely fluid stars in general-relativistic
MHD with more realistic field geometries also showed oscillations near
the continuum edges
\citep{sotani_etal08a,sotani_etal08b,cerda_etal09,cbk09,csf09}. 
\citet{vl11} have
shown that if a natural frequency of the free crust happens to fall
within a gap, there is little loss of the crust's oscillation energy to the
core by resonant absorption; see also work by
\citet{ck11}. Frequency drift, due to the effects of
the continuum, generally occurs \citep{vl11}. This basic picture has
been further developed and refined 
\citep{gabler_etal11,ck11,gabler_etal12,vl12,pl13,gabler_etal13,gabler_etal_sf13,gabler_etal14}.
In particular, \citet{ck11} find that all observed QPOs in SGRs 1806
and 1900 can be accommodated with a particular arrangement of
continuum gaps.
\citet{ck12} find in a spherical model with a magnetic field with both poloidal
and toroidal components that peaks below about 100 Hz can persist in
the power spectrum, though these peaks are far broader than
observed. \citet{gabler_etal13}, \citet{gabler_etal_sf13},
\citet{pl13}, and
\citet{pl14} find a small number of low-frequency QPOs that exist in
the continuum gaps and at turning points for their assumed field
geometries that agree qualitatively with observations. Superfluidity
widen the gaps in the Alfv\'en continuum
\citep{gabler_etal_sf13,pl14}.

Most of the work on the global oscillation problem described above
relies on tuning the locations of the gaps and turning points in the
Alfv\'en continuum to accommodate the observed QPOs.  Here we take a
rather different view than in previous work, and assert that the
Alfv\'en continuum that has been so problematic is unlikely to exist
at all. The Alfv\'en continuum results from assuming {\em smooth},
poloidal field configurations. Given the convective dynamo that is
expected to operate in a proto-magnetar, and subsequent evolution of
the initial field, we expect the field to have high-order evolving
multipoles. We conjecture that the field is complex and tangled over
length scales small compared to the stellar radius, and smooth only on
average. High-order multipoles (the tangle), could be long lived in
the core where the electrical conductivity is high. \citet{vl11} have
argued that a highly-tangled field is likely to reduce or eliminate
the importance of the Alfv\'en continuum and to give an effective
shear modulus to the core fluid; they demonstrated that the continuum
is broken for a ``box'' neutron star.

Much work on the QPO problem has included realistic neutron structure,
specific magnetic field geometries, and the effects of general
relativity. Given that the interior field geometry will likely remain
unknown, we take a step back in sophistication and propose a
minimal, illustrative model of the low-frequency magnetar QPOs ($<160$
Hz). The star is essentially a self-gravitating, constant-density,
magnetized fluid whose oscillation frequencies are determined by the
anisotropic magnetic stresses from the large-scale organized field
(primarily the dipole component) and the approximately isotropic
magnetic stresses introduced by the tangled field; we show that
density stratification and the crust typically change the oscillation
frequencies by only $\sim 10$\% (though 30-50\% for some fundamental 
modes), so this simple model is an adequate
first approximation for
developing a quantitative understanding of torsional modes, while
elucidating the essential physics. General
relativity is included only as a redshift factor that reduces the
oscillation frequencies observed at infinity by about 20\%. We show
that if the energy density in the tangled field is comparable to that
in the large-scale field, a discrete normal mode spectrum beginning at
around 20 Hz with a spacing of 10 Hz appears naturally for axial
modes, generally consistent with observations of QPOs below 160 Hz,
and that all QPOs observed in SGRs 1806 and 1900 in this frequency
range can be accounted for to within 3 Hz. We do not address the
character of high-frequency modes due to limitations to our model that
we describe. To illustrate the key features of the model, the
normal-mode analysis will be restricted entirely to axial modes. (We
use the terms ``axial'' and ``torsional'' synonymously).

In \S \ref{eom} we derive the equations of motion for a star with a
tangled field, and show how the tangled field gives the core fluid an
effective shear modulus. In \S \ref{continuum}, we briefly review how
an Alfv\'en continuum arises for a uniform field. 
In \S \ref{isotropic_model}, we present
analytic solutions for a star with only a tangled field; these
solutions are useful for understanding the results of a more general
field configuration consisting of an organized field plus a tangled
field, described in \S \ref{anisotropic_model}. In \S
\ref{stratification}, we show that realistic stellar structure and
crust stresses both have relatively small effects on most of the
eigenfrequencies for torsional modes if the energy density in the
tangled field is comparable to or larger than that in the organized
field. In \S
\ref{energetics}, we study the energetics of QPO excitation. 
In \S \ref{conclusions}, we discuss the implications of our work,
our key results, limitations of the model, and future
improvements. In Appendix \ref{sov}, we describe the variable
separation procedure we used to obtain the solutions of \S
\ref{anisotropic_model}. In Appendix
\ref{crust}, we use a simple model to show why the crust has a
negligible effect.

\section{Equations of Motion}

\label{eom}

The matter is subject to magnetic and material stresses. We will
find that QPOs in SGRs 1806 and 1900 can be explained by average
magnetic fields above $10^{15}$ G, comparable to the upper critical
field for superconductivity, so we take the star to be a normal
conductor. The dipole field could be much smaller than the average
field, and will not affect the basic picture provided the protons are
normal; we do not consider the effects of superconducting protons in
this preliminary study. 

The stress tensor for
matter permeated by a field $\Bbf$ is 
\begin{equation}
T_{ij}=\frac{1}{4\pi}\left[B_iB_j-\frac{1}{2}B^2\delta_{ij}\right]+
\mu\left(\nabla_i u_j +\nabla_j u_i\right).
\label{T}
\end{equation}
where $\mu$ is the shear modulus of the crust. We assume a perfect
conductor, so that perturbations in the field satisfy
\begin{equation}
\delta\Bbf=\nabla\times(\ubf\times\Bbf), 
\label{flux_freezing}
\end{equation}
where $\ubf$ is the displacement vector of a mass element. We
specialize to shear perturbations, so that $\nabla\cdot\ubf=0$, for
which eq. (\ref{flux_freezing}) becomes
\begin{equation}
\delta\Bbf=(\Bbf\cdot\nabla)\ubf-(\ubf\cdot\nabla)\Bbf.
\end{equation}
For a displacement $\ubf$, the stress tensor is perturbed by
\begin{equation}
\delta T_{ij}=\frac{1}{4\pi}\left[
B_jB_k\nabla_k u_i 
-B_ju_k\nabla_k B_i 
-\frac{1}{2}\delta_{ij}B_kB_l\nabla_lu_k
+\frac{1}{2}\delta_{ij}B_k u_l\nabla_l B_k\right] 
+\mu\nabla_iu_j+\mbox{ transpose}, 
\label{dT}
\end{equation}
where repeated indices are summed and $\mu$ is the shear modulus of
the matter, non-zero only in the crust, and $B_i$ denotes a component
of the unperturbed field. We include $\mu$ here only for completeness;
we will ultimately find that material stresses are dominated by those
from the tangled field.

We treat the magnetic field as consisting of an organized, 
largely dipolar contribution $\Bbf_o$, plus a much more complicated
tangled component $\Bbf_t$:
\begin{equation}
\Bbf=\Bbf_o+\Bbf_t.
\end{equation}
We assume that the field is tangled for length scales smaller than
$l_t$, small compared to the stellar radius. Given the uncertainties
in the overall field structure, we take $\Bbf_o$ constant for
simplicity. We denote volume averages over $l_t^3$ as $\ev{...}$. We assume
that different components of the tangled field are uncorrelated on
average. Under this assumption, the  tangled field can contribute only
isotropic stress over length scales above $l_t$, so that
\begin{equation}
\ev{B_iB_j}=B^o_iB^o_j +\ev{B_t^2}\delta_{ij},
\label{dc1}
\end{equation}
where $\ev{B_t^2}$ is a constant. 

To treat the tangled field's contribution to the stress, we average
the perturbed stress tensor of eq. (\ref{dT}).  
Eqs. (\ref{dT}) and (\ref{dc1}) to obtain
\begin{equation}
\langle\delta T_{ij}\rangle=
\frac{1}{4\pi}\left[
\ev{B_jB_k}\nabla_k u_i -u_k\ev{B_j\nabla_kB_i}
-\frac{1}{2}\delta_{ij}\ev{B_kB_l}\nabla_l u_k +
\frac{1}{2}\delta_{ij}u_l\ev{B_k\nabla_lB_k}\right]
+\mu\nabla_i u_j + \mbox{ transpose}, 
\label{dT1}
\end{equation}
where $u_i$ now denotes a component of the displacement field averaged
over $l_t^3$.

If different components of the tangled field are uncorrelated over
$l_t^3$, one component will also be uncorrelated with the gradient
of a different component, that is, 
\begin{equation}
\ev{B_i\nabla_k B_j}=\ev{(B^o_i+B^t_i)\nabla_k (B^o_j+B^t_j)}=
\ev{B^t_i\nabla_k B^t_j}=0 \qquad i\ne j.
\label{dc2}
\end{equation}
Since the tangled field varies over length scales smaller than $l_t$, 
a component of the tangled field will also be uncorrelated with the
gradient of the same component, so that
\begin{equation}
\ev{B^t_i\nabla_k B^t_i}=0, 
\label{dc3}
\end{equation}
which applies component by component, and therefore also in
summation, as given above. 
Using eqs. (\ref{dc1}), (\ref{dc2}), (\ref{dc3}), and
$\nabla\cdot\ubf=0$, eq. (\ref{dT1}) becomes
\begin{equation}
\ev{\delta T_{ij}}=
\frac{1}{4\pi}\left(
B_j^o B_k^o\nabla_k u_i + B_i^o B_k^o\nabla_k u_j
-\delta_{ij}B_k^o B_l^o \nabla_l u_k\right)
+\left(\frac{1}{4\pi}\ev{B_t^2}+\mu\right)
\left(\nabla_i u_j +\nabla_j u_i\right).
\label{dT_final}
\end{equation}
The tangled field gives the fluid an effective shear modulus of
$\ev{B_t^2}/4\pi$, and enhances the rigidity of the solid.

Upon comparing our mode calculations with data, we will find that the
total energy in shear waves in the core greatly dominates that in
the crust. We henceforth ignore crust rigidity, and justify this
approximation in \S \ref{anisotropic_model} and Appendix \ref{crust}. 

We will be interested in modes with wavelengths greater than $l_t$, for
which the equation of motion is 
\begin{equation}
\rho_d\frac{\partial^2 u_j}{\partial t^2}=\nabla_i \ev{\delta T_{ij}}, 
\label{eom_gen}
\end{equation}
where $\rho_d$ is the dynamical mass density of matter that is frozen
to the magnetic field. If the the protons are normal, as we have
assumed, there will no entrainment between the protons and neutrons;
entrainment of the neutrons is in any case a small effect if the both
the protons and neutrons are superfluid \citep{ch06}.  We take
$\rho_d$ to be the proton density $x_p\rho$, where $\rho$ is the mass
density, and $x_p\sim 0.1$ is the average proton mass fraction in the
core.

We neglect coupling of the stellar surface to
the magnetosphere, and treat the surface as a free
boundary with zero traction, thus ignoring momentum flow into the
magnetosphere. Under this assumption, the traction at the surface vanishes:
\begin{equation}
\hat{r}_i\ev{\delta T_{ij}}=0,
\label{eom1}
\end{equation}
where $\hat{r}$ is the unit vector normal to the stellar surface. 

Given the uncertainties in the field geometry, we henceforth consider
a uniform star of density $\rho=3M/4\pi R^3$ for illustration, where
$M$ and $R$ are the stellar mass and radius, and take
$\Bbf_o=\hat{z}B_o$ where $B_o$ is constant.  Eqs. (\ref{dT_final})
and (\ref{eom_gen}) give
\begin{equation}
c_o^2\frac{d^2\ubf}{dz^2}
-c_o^2\nabla\frac{d u_z}{dz}+
c_t^2\nabla^2\ubf+\omega^2 \ubf=0,
\label{eom_vector}
\end{equation}
where $c_o^2\equiv B_o^2/(4\pi\rho_d)$ and $c_t^2\equiv
\ev{B_t^2}/(4\pi\rho_d)$; $c_o$ is the speed of Alfv\'en waves
supported by the organized field, and $c_t$ is the speed of transverse
waves supported by the isotropic stress of the tangled field. We show
in \S \ref{stratification} that realistic stellar structure changes
the eigenfrequencies of torsional modes by typically $\sim 10$\%
(but by up to $\sim 50\%$ for the lowest-frequency fundamentals), so
that a constant-density model is a good first approximation.

\section{The Alfv\'en Continuum}

\label{continuum}

For finite $\Bbf_o$ and $\Bbf_t=0$, there exists a
continuum of axial modes \citep{levin07}, 
given by eq. (\ref{eom_vector}) 
\begin{equation}
c_o^2\frac{d^2\ubf}{dz^2}-c_o^2\nabla\frac{d u_z}{dz}+\omega^2\ubf=0, 
\end{equation}
In cylindrical coordinates ($s,\phi,z$), axial modes are given by
$\ubf=u_\phi\,\hat{\phi}$. 
For a constant field, and no crust, field
lines have a continuous range of lengths between zero and $2R$,
determined by the cylindrical radius $s$. 
Within the approximation of ideal
MHD, fluid elements at different cylindrical radii cannot exchange
momentum. At a given $s$, the length of a field line is
$2\sqrt{R^2-s^2}$. The 
solutions have even parity ($\cos kz$) and odd parity ($\sin kz$).  The
requirement that the traction vanish at the stellar surface gives the
spectrum
\begin{equation} \label{cont_even}
\nu_n=\frac{n}{2}\frac{c_o}{\sqrt{R^2-s^2}} 
\qquad\mbox{even parity}
\end{equation}
\begin{equation} \label{cont_odd}
\nu_n=\frac{(2 n+1)}{4}\frac{c_o}{\sqrt{R^2-s^2}}  \qquad\mbox{odd parity.}
\end{equation}
where $n$ is an integer, beginning at zero for the odd-parity
modes, and $\nu=\omega/2\pi$.  Because $s$ is a
continuous variable, for every $n$ there is a continuous spectrum of
modes for this simple magnetic geometry.  An infinite sequence of
continua begins at $\sim 7 (2n+1) \,B^o_{15}$ Hz for odd-parity modes
and $ \sim 7 (2 n)
\,B^o_{15}$ Hz for the even-parity modes, where $B^o_{15}\equiv
B_o/(10^{15}\mbox{ G})$.  The full spectrum begins at $\sim
7\,B^o_{15}$ Hz; there is also a zero-frequency mode corresponding to
rigid-body rotation.  The same conclusion holds for more general
axisymmetric field geometries, though certain geometries give gaps in
the continuum.

\section{Isotropic Model}

\label{isotropic_model}

We now turn to the opposite extreme of a tangled field that dominates
the stresses, taking $\Bbf_o=0$, and solving the resulting isotropic
problem. This problem provides useful insight into the mode structure
of the more general problem with non-zero $\Bbf_o$ and $\ev{B_t^2}$. 
For this case, eq. (\ref{eom_vector}) becomes
\begin{equation}
c_t^2\nabla^2\ubf+\omega^2\ubf=0, 
\end{equation}
Subject to the restriction $\nabla\cdot\ubf=0$, the solutions for
spheroidal modes ($u_r=0$), can be
separated as
\begin{equation}
u_\phi=w(r)\frac{\partial}{\partial\theta} Y_{lm}(\theta,\phi){\rm
e}^{i\omega t}
\end{equation}
\begin{equation}
u_\theta=-w(r)\frac{1}{\sin\theta}\frac{\partial}{\partial\phi} 
Y_{lm}(\theta,\phi){\rm
e}^{i\omega t}. 
\end{equation}
The radial function $w(r)$ satisfies Bessel's equation: 
\begin{equation}
\left(\frac{d^2}{dr^2}+\frac{2}{r}\frac{d}{dr}-
\frac{l(l+1)}{r^2}+k^2\right)w(r)=0,
\label{radial_equation}
\end{equation}
where $k\equiv\omega/c_t$. 

The solutions that are bounded at $r=0$ are the spherical Bessel
functions $j_l(kr)$. 
Zero traction at the stellar surface gives
\begin{equation}
\left[\frac{dj_l}{dr}-\frac{j_l}{r}\right]_{r=R}=0 
\label{bc_isotropic}
\end{equation}

For each value of $l$, eq. (\ref{bc_isotropic}) has solutions
$x_{l,n}\equiv k_{l,n} R$, where $n=0,1,2...$, the overtone number,
gives the number of nodes in $j_l(r)$. The eigenfrequencies are
\begin{equation}
\omega_{l,n}=\left(1-\frac{R_s(M)}{R}\right)^{1/2}
\left(\frac{\ev{B_t^2}R}{3x_pM}\right)^{1/2} x_{l,n}
\end{equation}
where a redshift factor $z\equiv(1-R_s(M)/R)^{1/2}$ has been
introduced; $R_s$ is the Schwarzchild radius. In terms of fiducial
values
\begin{equation}
\nu_{l,n}(\mbox{Hz})=\frac{\omega_{l,n}}{2\pi}=
4.3\, \left(\frac{z}{0.77}\right)
\left(\frac{R}{10\mbox{ km}}\right)^{1/2}
\left(\frac{M}{1.4M_\odot}\right)^{-1/2}
\left(\frac{x_p}{0.1}\right)^{-1/2}
\left(\frac{\ev{B_t^2}^{1/2}}{10^{15}\mbox{ G}}\right)
\, x_{l,n}\mbox{ Hz}.
\label{nu_isotropic}
\end{equation}

In Table \ref{nu_isotropic_values}, we give some of the eigenfrequencies for
these fiducial values. We note that for $\ev{B_t^2}^{1/2}=10^{15}$ G,
this simple model gives fundamental frequencies below 20 Hz,
with a spacing of about 15 Hz for the overtones. 

\begin{table}
\begin{tabular}{l|lllllll}
\hline
$l$ & $n=0$ & $n=1$ & $n=2$ & $n=3$ & $n=4$ & $n=5$ & $n=6$\\
\hline
1 & 0 & 24 & 39 & 52 & 66 & 79 & 93\\
2 & 11 & 30 & 45 & 58 & 72 & 85 & 99\\
3 & 16 & 36 & 50 & 64 & 78 & 92 & 105 \\
4 & 22 & 41 & 56 & 70 & 84 & 98 & 111\\
5 & 27 & 46 & 61 & 76 & 90 & 103 & 117\\
6 & 31 & 52 & 67 & 81 & 95 & 109 & 123\\
\end{tabular}
\caption{Eigenfrequencies in Hz for the fiducial values of
eq. (\ref{nu_isotropic}). }
\label{nu_isotropic_values}
\end{table}

For $l=1$, eq. (\ref{bc_isotropic}) has a solution $w(r)=r$ for
$k=0$.  This
solution corresponds to rigid-body rotation and we label it
$n=0$. This solution is of no physical significance to the mode
problem we are addressing, but we include it in Tables 1-3 for
completeness.

Frequencies above 160 Hz correspond to $k_{n,l}R\gap 30$, or
wavelengths $\lap 0.2R$. At these high wavenumbers, the wavelength of
the mode could be comparable to or smaller than the length scale $l_t$
over which the field can be considered tangled, in which case the
averaging procedure of \S \ref{eom} would break down. In this case,
the mode frequencies will be determined, at least in part, by the
detailed (and unknown) field geometry. Henceforth, we restrict the
analysis to $\nu<160$ Hz, the range in which the low-frequency QPOs lie.

\section{Anisotropic Model}

\label{anisotropic_model}

We now turn to the general problem of non-zero $\Bbf_o$ and
$\ev{B_t^2}$, and show that even a small amount of stress from the
tangled field breaks the Aflv\'en continuum very effectively. The
normal modes of the system are similar to those found in the isotropic
problem for $\ev{B_t^2}\simeq B_o^2$. 

The modes are given by eq. (\ref{eom_vector}). 
This vector equation, subject to boundary conditions on the surface of a
sphere, is difficult to solve in general. For illustration, we
specialize to axial modes, $u_\theta=u_r=0$, giving in cylindrical 
coordinates $(s,\phi,z)$:
\begin{equation}
c_o^2\frac{d^2u_\phi}{dz^2}+
c_t^2\nabla^2 u_\phi-c_t^2\frac{u_\phi}{s^2}+\omega^2 u_\phi=0. 
\end{equation}
The third term results from derivatives of $\hat{\phi}$ in
the original vector equation (\ref{eom_vector}).
Defining 
$b_t^2\equiv \ev{B_t^2}/B_o^2$, the ratio of the energy
density in the tangled field to that in the organized field, the above
equation becomes
\begin{equation}
c_o^2\left[b_t^2\frac{1}{s}\frac{\partial}{\partial s}
\left( s\frac{\partial u_\phi}{\partial s}\right)
+\left(1+b_t^2\right)\frac{\partial^2 u_\phi}{\partial z^2} 
-b_t^2\frac{u_\phi}{s^2}\right]+\omega^2 u_\phi=0\,.
\label{eomt}
\end{equation}
The zero-traction boundary condition (eq. \ref{eom1})
at the stellar surface is
\begin{equation}
b_t^2\left( s \frac{\partial u_\phi}{\partial s} -u_{\phi} \right)
+\left(1+b_t^2 \right)z\frac{\partial u_\phi}{\partial z}=0.
\label{bct}
\end{equation}
Equation (\ref{eomt}) is solved in the domain $z\geq 0$, and so at
$z=0$ we require
\begin{equation}
  u_\phi =  0\, ,
\end{equation}
for modes with {\it odd} parity about $z=0$, and 
\begin{equation}
  \frac{\partial u_\phi}{\partial z} =  0\, ,
\end{equation}
for modes with {\it even} parity.
The system is separable with a coordinate transformation; the details
are given in Appendix \ref{sov}. We solve eqs. (\ref{sepu}),
(\ref{sepv}), and (\ref{bcu}) numerically to obtain the normal mode
frequencies. The solutions are given by two quantum numbers: $\kappa$
and the overtone number $n$. $\kappa$ maps smoothly to $l(l+1)$ in the
limit $c_o^2\rightarrow 0$, so we use $l$ and $n$ for convenience in
labeling the modes. We refer to $n=0$ for a given $l$ as the
fundamental for that value of $l$.

\begin{figure*}
\centering
\includegraphics[width=.4\linewidth]{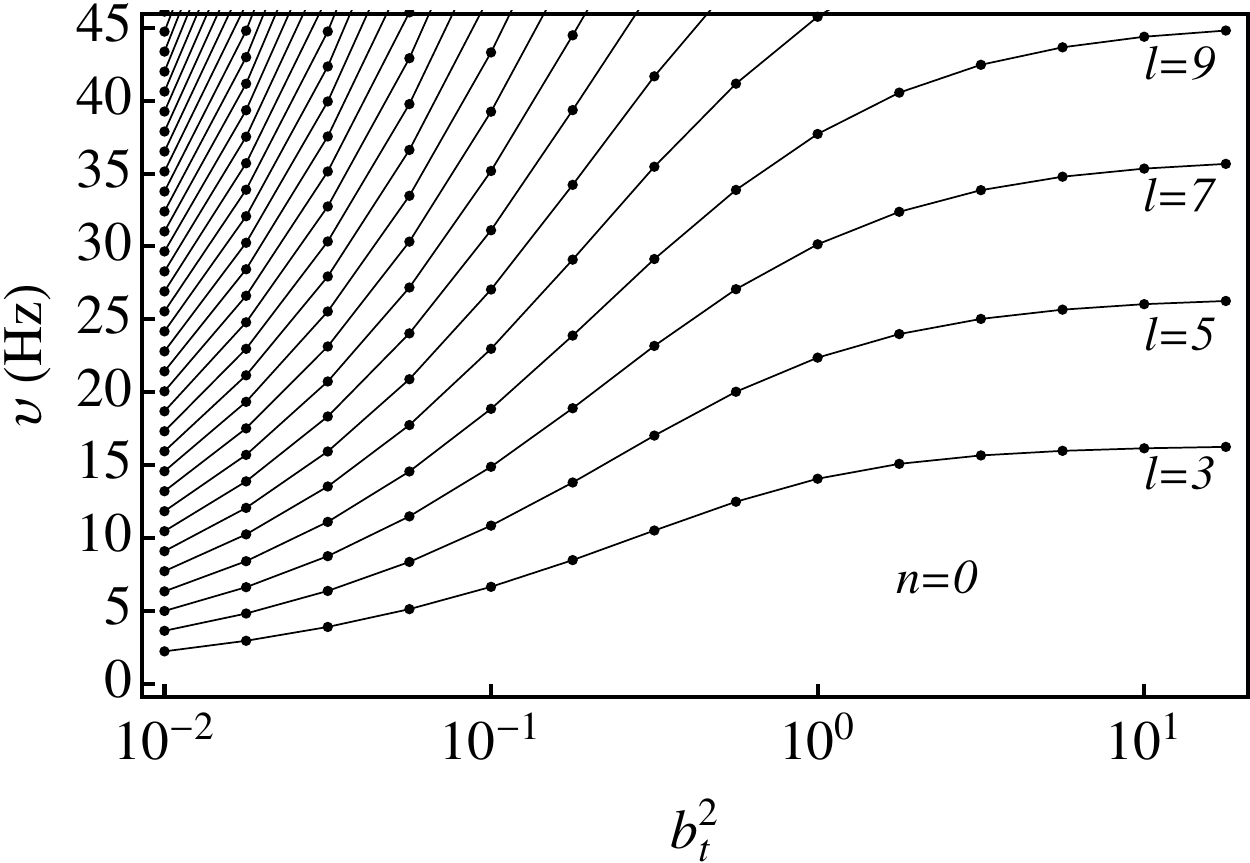} 
\includegraphics[width=.4\linewidth]{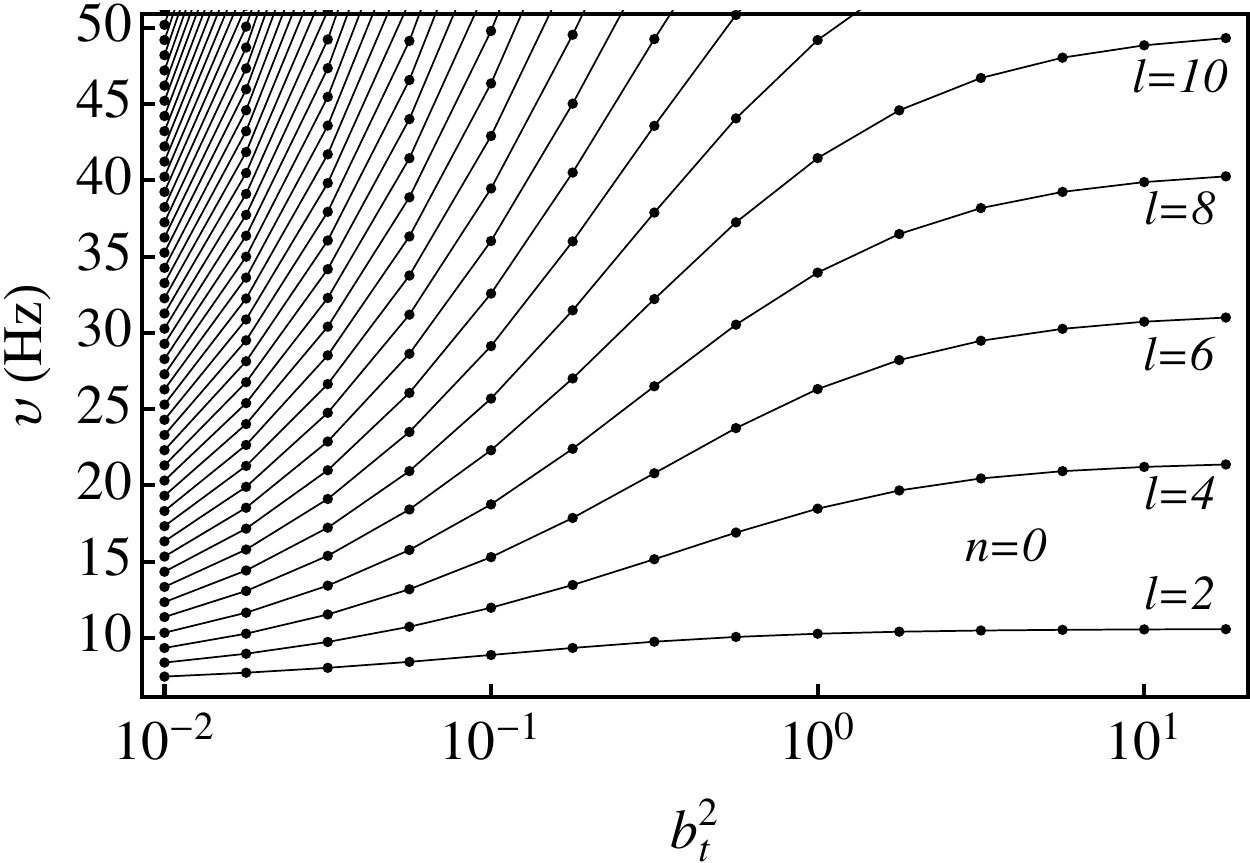} \\
\includegraphics[width=.4\linewidth]{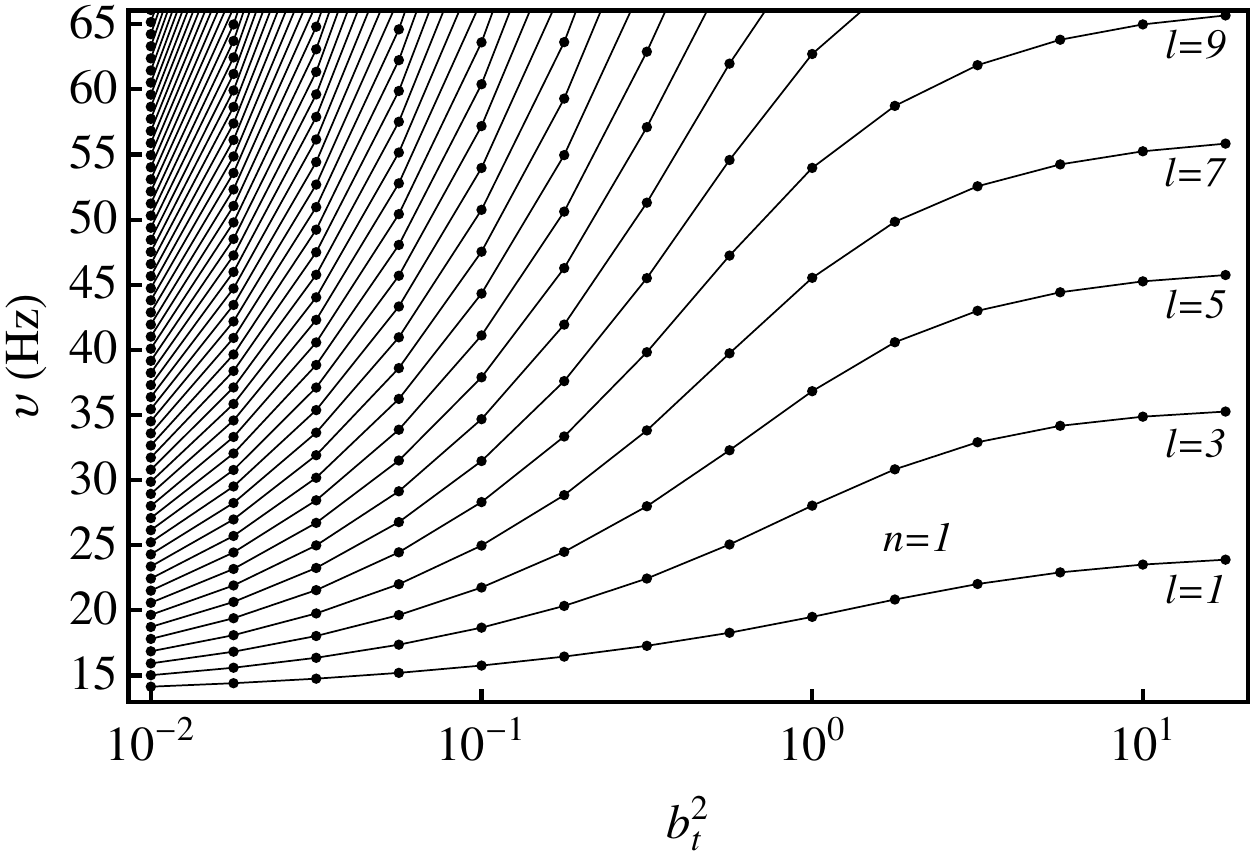} 
\includegraphics[width=.4\linewidth]{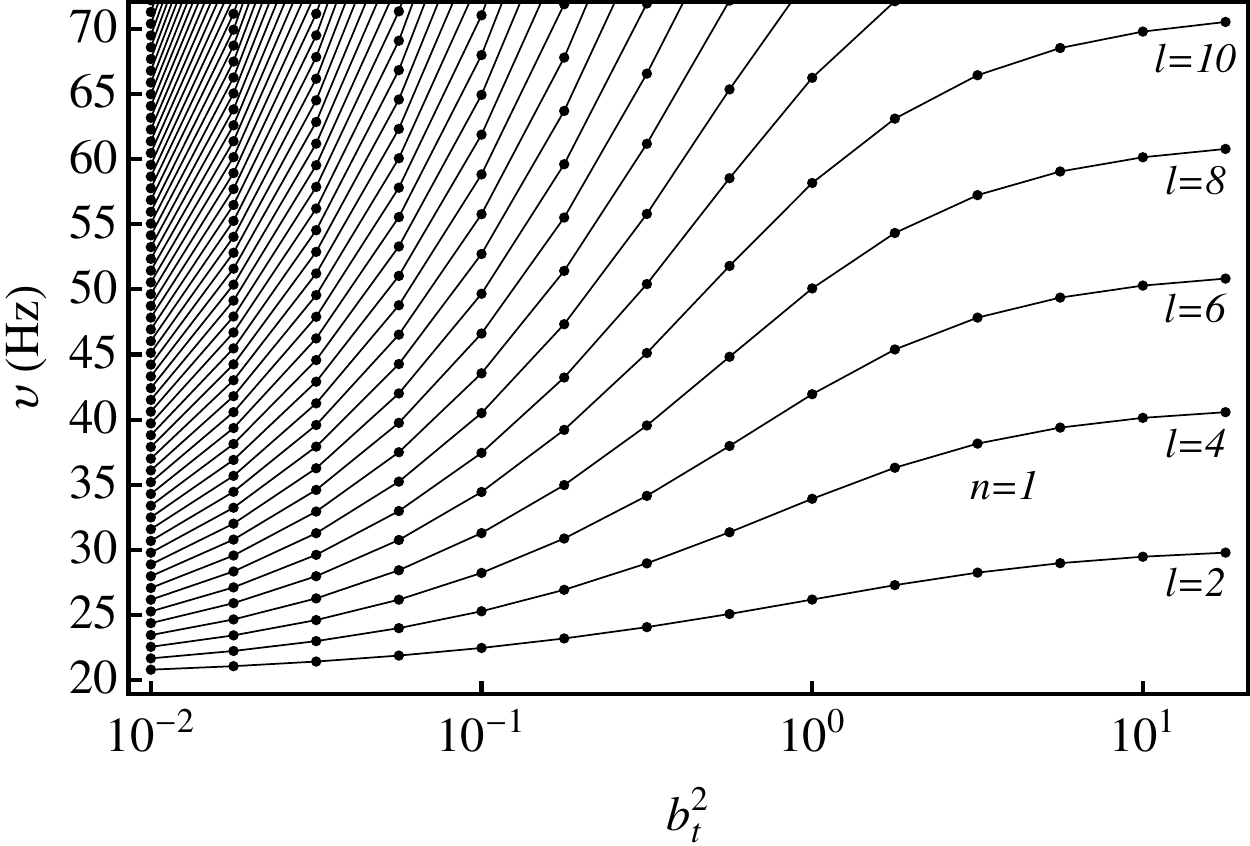}
\caption{Splitting of the Alfv\'en continuum as a tangled field is
added to a smooth field. We have fixed
$(B_0^2+\ev{B_t^2})^{1/2}=B_0(1+b_t^2)^{1/2}$ to $10^{15}$ G to
illustrate the smooth transition from the continuum to the isotropic
tangle. For $l$ odd and $n=0$ (upper left panel), low-frequency
modes exist that go to zero frequency for $b_t^2\rightarrow 0$; see
equation (\ref{l_0_scaling}).}
\label{splitting}
\end{figure*}

The mode structure is shown in Figure \ref{splitting} for constant {\it
total} magnetic energy, with $(B_0^2+\ev{B_t^2})^{1/2}$ fixed at
$10^{15}$ G. 
Modes for which $u_\phi$ has even parity (odd $l$) and odd parity (even $l$) about $z=0$ have been plotted separately.
For large $b_t^2$ the isotropic solution presented in Table
\ref{nu_isotropic_values} is recovered.  As $b_t^2$ is reduced, the
modes become more closely spaced, approaching a continuum as
$b_t^2\rightarrow 0$. In the limit $b_t^2\rightarrow 0$ the sequence
of continua begins at $\sim 7(2n+1)$ Hz for odd-parity modes (even
$l$) and $\sim 14 n$ Hz for even-parity modes (odd $l$), in agreement
with the continuum sequences described by eqs. (\ref{cont_even}) and
(\ref{cont_odd}).  The $n=0$ odd modes approach zero as $b_t^2
\rightarrow 0$ and scale as 
\begin{equation}
\nu=
6.70\, l \, b_t \, \left(\frac{z}{0.77}\right)
\left(\frac{R}{10\mbox{ km}}\right)^{1/2}
\left(\frac{M}{1.4M_\odot}\right)^{-1/2}
\left(\frac{x_p}{0.1}\right)^{-1/2} 
\left(\frac{B_o}{10^{15}\mbox{ G}}\right)
\, \mbox{ Hz}.
\label{l_0_scaling}
\end{equation}
These modes approach rigid-body rotation solutions of the continuum
coupled together by the tangled field in the $b_t\rightarrow 0$
limit.
Examination of the eigenmodes as $b_t^2\rightarrow 0$ also shows that the oscillation amplitude becomes sharply peaked at a specific value of the cylindrical radius
$s$ and vanishes everywhere else, in agreement with the continuum
solution in \S\ref{continuum}. 

\begin{figure*}
\centering
\includegraphics[width=.48\linewidth]{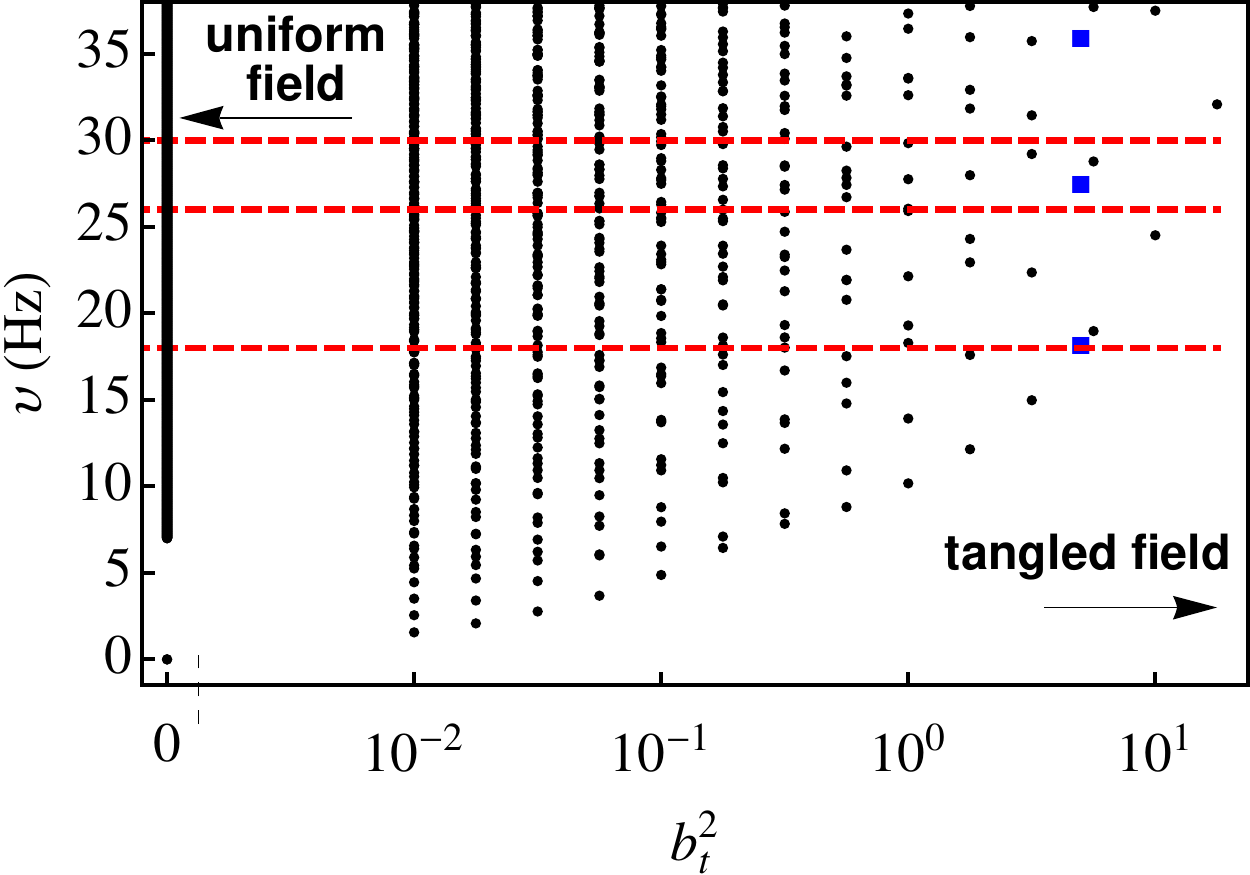} 
\includegraphics[width=.49\linewidth]{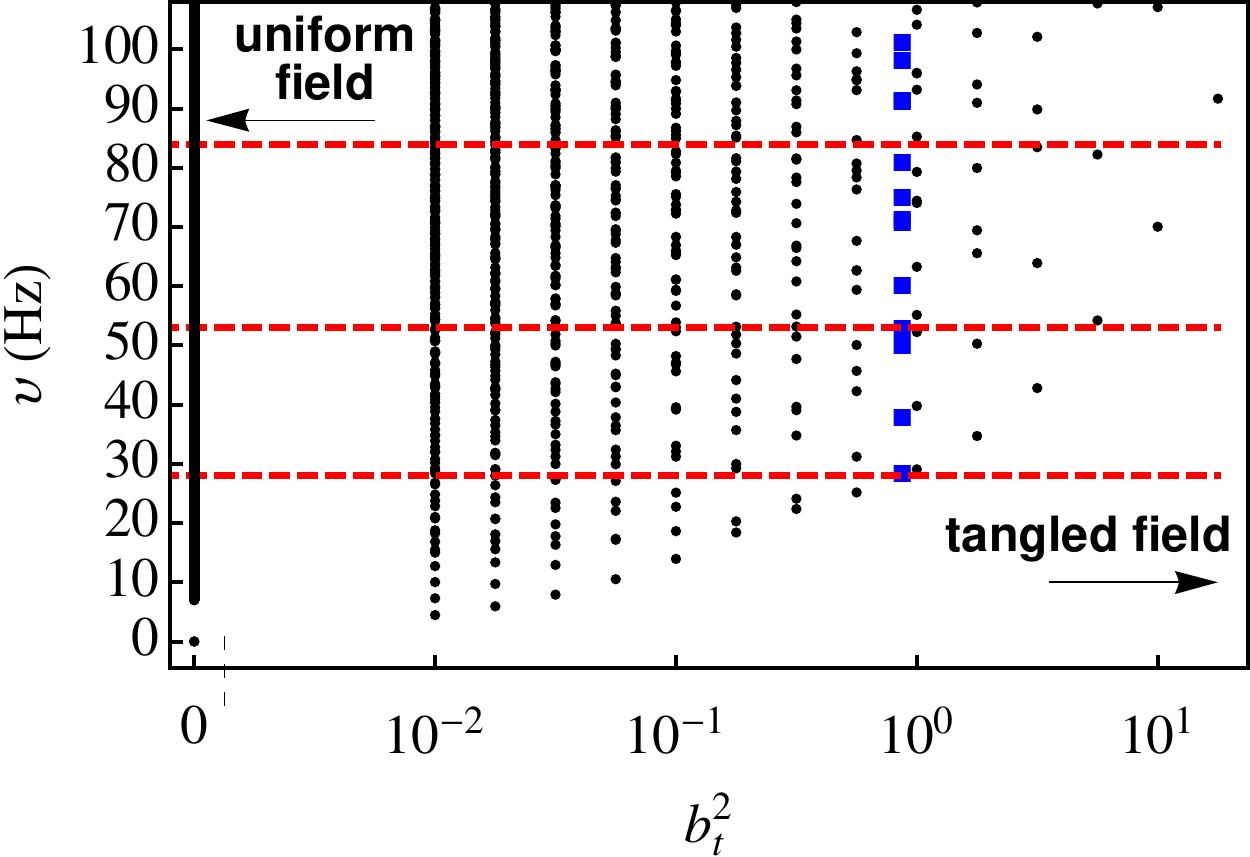} 
\caption{Splitting of the Alfv\'en continuum (the column on the far
left of each figure) as a tangled field is
added to a smooth field in SRG 1806 (left) and 1900 (right).  The
ordered field is fixed at the dipole spin-down value: $2\times
10^{14}$ for SGR 1806 and $2 \times 10^{15}$ for SGR 1900. Red-dashed
lines correspond to observed QPO frequencies. Solid-blue squares
represent the value of $b_t^2$ that best fits the data in each object,
though note that only three observed QPOs are shown on the scales of
these plots. The lowest sequence of points approach the $k=0$
rigid-body mode (not shown here), according to eq. (\ref{l_0_scaling}) .}
\label{dots}
\end{figure*}

The splitting of the Alfv\'en continuum is shown in a different way in
Figure \ref{dots}, where realistic predictions are made for the
eigenfrequencies in SGR 1806 (left) and 1900 (right).  In contrast
with Figure \ref{splitting} where the {\it total} magnetic field was
held constant ($(B_0^2+\ev{B_t^2})^{1/2}=10^{15}$ G), we now fix the
{\it ordered} field $B_o$ in each magnetar to its observed dipole
spin-down value: $7\times 10^{14}$ G for SGR 1806 and $2\times
10^{15}$ G for SGR 1900.  We take the fiducial values of
$M=1.4M_\odot$, $R=10$ in each magnetar.  Eigenfrequencies for all $n$
and $l$ for the range of frequencies shown are plotted.  For
$b_t^2=0$, there exists an Alfv\'en continuum that begins at $\simeq
7\,B^o_{15}$ Hz, with a zero-frequency solution corresponding to
rigid-body rotation. As $b_t^2$ is increased, the continuum splits
into discrete normal modes.  As $b_t^2$ is further increased and the
modes spread out further, the higher-frequency modes move off the
diagram.  For large $b_t^2$ the eigenfrequencies become large due to
the large tangled field.

The observed QPO frequencies in SGRs 1806 and 1900 are shown in 
Figure \ref{dots} as horizontal red dashed lines.  For given $M$ and $R$, the
model has only $b_t^2$ as a free parameter, and gives quantitative
predictions.  Assuming the lowest observed QPO corresponds to the
lowest eigenfrequency in the spectrum fixes the value of $b_t^2$ in
each object, giving 5.0 and 0.87 for SGR 1806s and 1900 respectively.
Eigenfrequencies for these $b_t^2$ are plotted as solid blue
squares.

\begin{table}
\begin{tabular}{l|lllllll}
\hline
$l$ & $n=0$ & $n=1$ & $n=2$ & $n=3$ & $n=4$ & $n=5$ & $n=6$\\
\hline
1 & 0 &   39 &   62 &   85 &   108 &  130 &   {\bf 153}\\
2 &   {\bf 18} & 49 &   73 &   96 &   119 &  142 &  \\
3 &   {\bf 27} &  {\bf 58} & 81 &  104 &   {\bf 126} &   {\bf 148} & \\
4 &   36 &   67 &   {\bf 91} &   114 &   137 &   160 & \\
5 &  44 & 76 &  100 &   123 &   145 &   & \\
6 &   52 &  84 &  109 &  133 &   155 &   & \\
7 &  {\bf 59} & {\bf 93} & 118  & 142   &    &   & \\
\end{tabular}
\caption{Eigenfrequencies in Hz for SGR 1806 with $B_o=7\times
10^{14}$ and $b_t^2=5.0$. Numbers in boldface are within 3 Hz of an
observed QPO, and represent plausible mode identifications. 
We do not list frequencies above 160 Hz.}
\label{nu_SGR1806_values}
\end{table}

\begin{table}
\begin{tabular}{l|llllll}
\hline
$l$ & $n=0$ & $n=1$ & $n=2$ & $n=3$ & \\
\hline
1 & 0 &  {\bf 52} & 89 & 126 \\
2 &  {\bf 28} & 71 &  108 & 144 \\
3 & 38 &  75 & 109 &  145   \\
4 & 50 &  91 & 127 &    \\
5 & 60 &  98 & 130 &    \\
6 &  70 & 112 & 147 &    \\
7 &  {\bf 81} & 121 & {\bf 152} &    \\
8 &  91 & 134 &  &    \\
9 &  101 & 144 &  &    \\
10 &  111 & {\bf 155} &  &    \\
\end{tabular}
\caption{Eigenfrequencies in Hz for SGR 1900 with $B_o=2\times
10^{15}$ and $b_t^2=0.87$. Numbers in boldface are within 3 Hz of an
observed QPO, and represent plausible mode identifications.
We do not list frequencies above 160 Hz.}
\label{nu_SGR1900_values}
\end{table}

In Tables \ref{nu_SGR1806_values} and \ref{nu_SGR1900_values}, we give
 frequencies for SGR 1806 and 1900 at $b_t^2=5.0$ and $0.87$ respectively.  
Numbers in boldface denote normal mode frequencies
that are within 3 Hz of an observed QPO in both SGRs 1806 and 1900, 
and represent plausible mode identifications. Within these
tolerances, {\em every observed QPO frequency below 160 Hz in SGRs 1806 and
1900 is accounted for}. For SGR 1806, the model does not give the distinct
26 Hz and 30 Hz that have been reported, but we note that these
features are quite broad \citep{ws06}. The reported frequencies 
have uncertainties, and drift with time, and analysis has not yet been
done to determine if the data favor two frequencies around 26-30 Hz,
or only one. In our model, both are accommodated by
the 27 Hz frequency for this example. If distinct QPOs at 26 and 30 Hz
could be ascertained, this would challenge our model.

\begin{figure*}
\centering
\includegraphics[width=.4\linewidth]{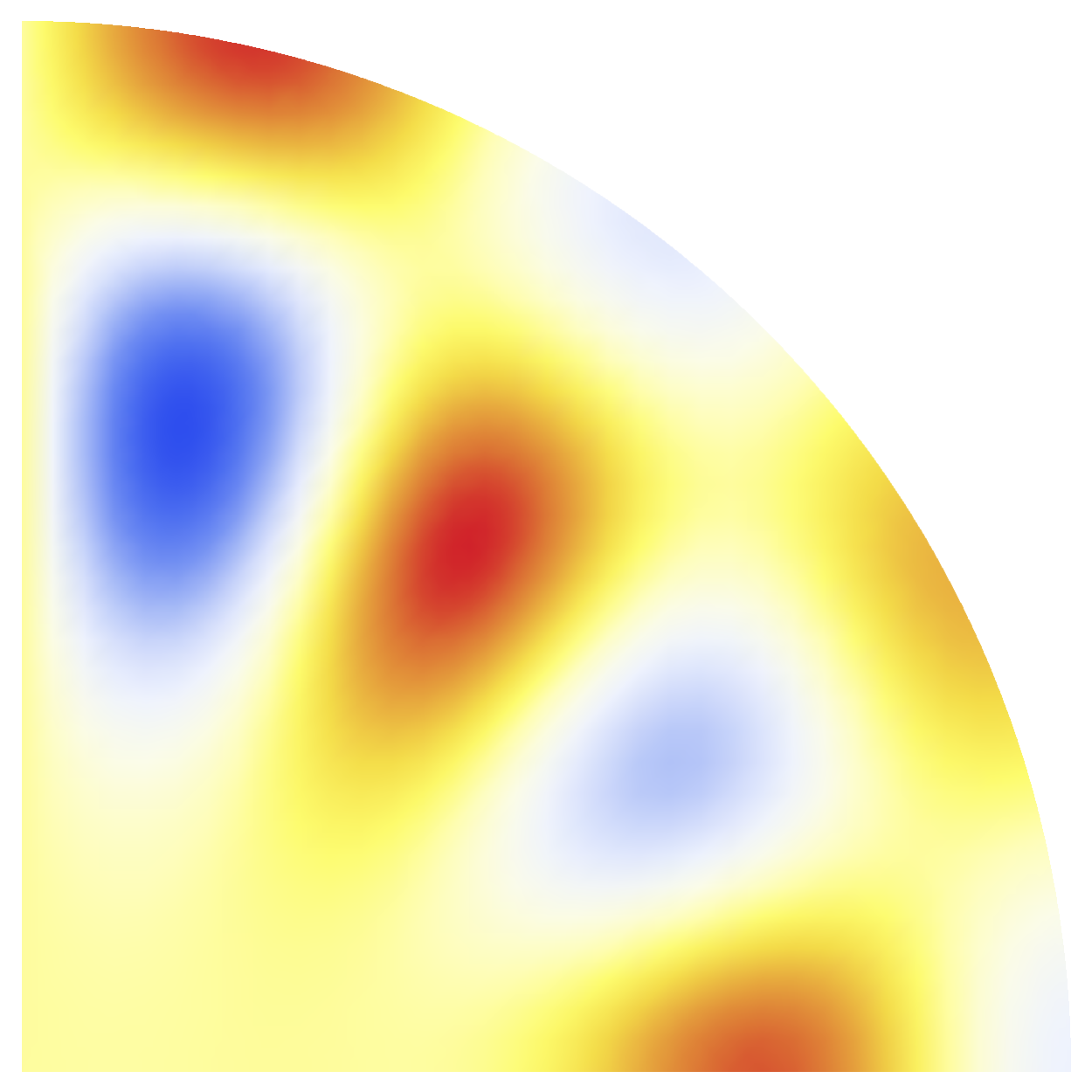} 
\includegraphics[width=.4\linewidth]{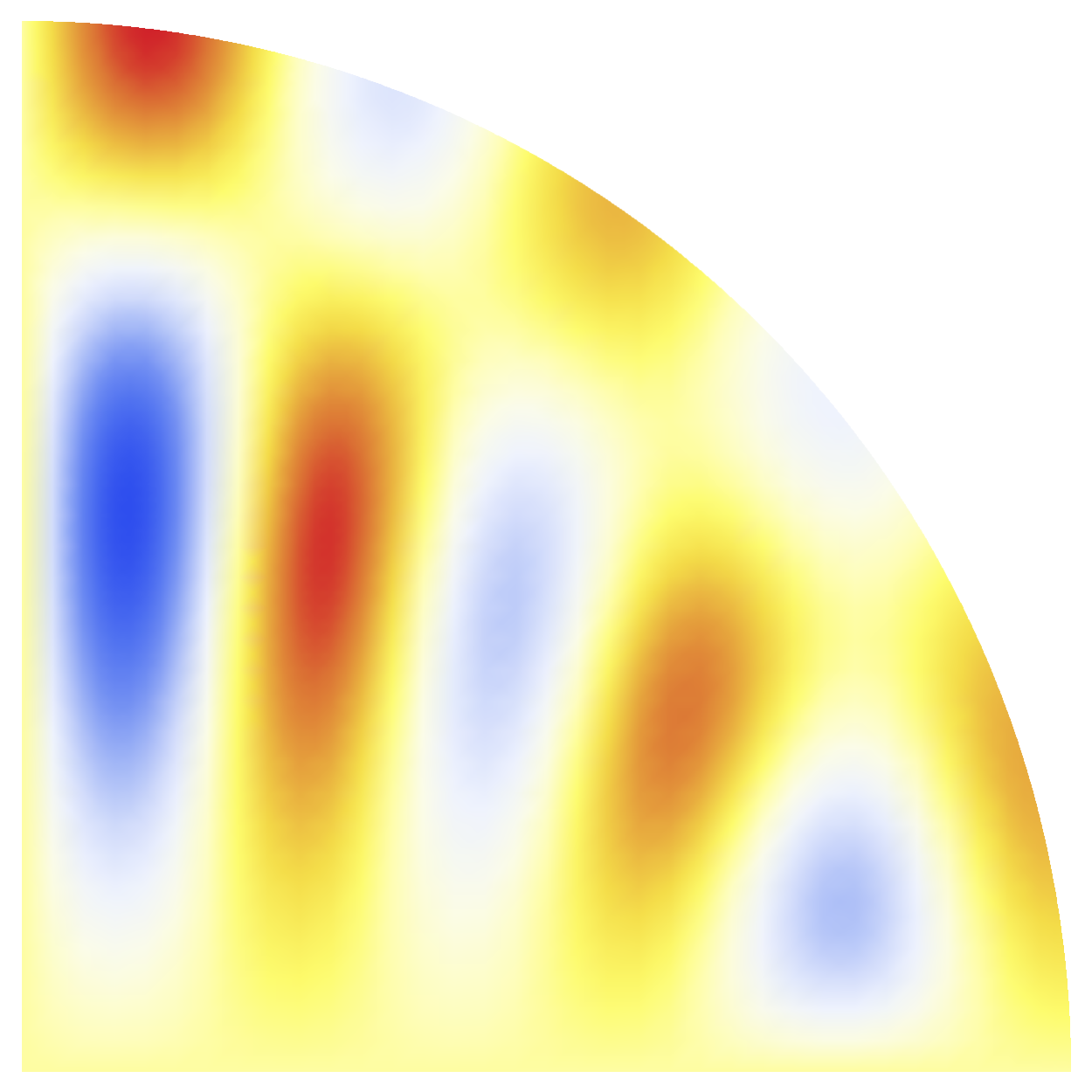} \\
\caption{Example eigenmodes for SGR 1806s (left) and 1900 (see text
for details). Blue denotes motion
out of the page for a given phase, red denotes into the page, and white
denotes zero amplitude.  The left panel has $b_t^2=5$ and exhibits the spherical structure of the isotropic limit, while the right has $b_t^2=0.86$ and exhibits the cylindrical structure of the continuum.}
\label{modes}
\end{figure*}

Examples of eigenmodes are plotted in Figure \ref{modes}.  The left
panel shows the 93 Hz mode of SGR 1806, with the attribution $l=7$,
$n=1$, while the right panel shows the 155 Hz of SGR 1806, with the
attribution $l=10$, $n=1$. The left panel has $b_t^2=5.0$, which is
close to the isotropic limit, and shows a large degree of spherical
structure symmetry.  In the plot on the right, $b_t^2=0.87$, and the
mode shows mostly cylindrical structure, a characteristic feature of
the continuum.

Though our simple, constant density model is intended mostly for
illustration, it accurately reproduces all observed QPOs in SGRs 1806
and 1900 though the spectrum is rather dense with a spacing of about
10 Hz. The main goal has been to show in a quantitative manner how the
Alfv\'en continuum is broken. As we show in \S \ref{stratification},
the effects of realistic stellar structure are generally small 
(except for low-$l$ fundamentals), and so this constant-density model
is sufficiently accurate for semi-quantitative study of torsional
oscillations. Many of the predicted frequencies have not been
observed, and we discuss this further below.

All examples shown here are for $R=10$ km. If $R$ is reduced, it is
necessary to decrease $b_t^2$ to obtain similar fits to the data. The
spectrum becomes generally less dense. Larger $R$ has the opposite
effect.

\section{Effects of stratification and the Crust}

\label{stratification}

We now show that density stratification and crust rigidity gives 
corrections to the eigenfrequencies of the constant-density model
considered thus far of typically $\sim 10$\%, but up to 50\% for
low-$l$ fundamentals. Because our best fits to the data for the
constant-density model give $b_t^2=5$ for SGR 1806 and $b_t^2\sim 1$
for SGR 1900, we ignore the organized field, and study the isotropic
problem corresponding to a strong magnetic tangle.

We construct a relativistic star using the analytic representation of the Brussels-Montreal equation of state derived by \citet{potekhin_etal13}. For a 1.4 $M_\odot$ neutron star, we obtain a radius of 13.1 km. We take 
$x_p=0.1$, and and fix the magnetic field at $\ev{B_t^2}=10^{15}$ G. The shear
modulus is from
\citet{strohmayer_etal91}. For this spherical problem, the variable
separation proceeds as in \S \ref{isotropic_model}. We solve
numerically for the radial eigenfunctions and eigenfrequencies.  As a
baseline for comparison, Table \ref{nu_isotropic_values} can be scaled
according to eq. (\ref{nu_isotropic}) to obtain the eigenfrequencies
for a constant-density star of radius 13.1 km. We assume that the
dynamical density in the crust is equal to the total mass density, so
that the entrainment of superfluid neutrons in the crust 
\citet{chamel05,chamel12} is effectively perfect. Entrainment, or lack
of entrainment, has little effect on our results.

To explore the individual effects of stratification and the crust, we
first examined a stratified star with no crust.  We find that
stratification lowers the frequencies for all modes with the exception
of the $n=0$ modes, which are raised; typical changes are $<20$\% for
$n=0$ and -15\% for $n>0$. This general trend was also found by
\citet{pl14}. The reason for the frequency decrease (except for $n=0$)
can be seen by noting that the frequency scales as $R^{1/2}$ in the
case of uniform density; see eq. (\ref{nu_isotropic}). The stratified
star has more mass concentrated in the central regions of the star,
making it behave effectively as a smaller star compared to a
constant-density star of the same radius.

Next, we considered a stratified star with a crust.  The addition of a
crust increases all frequencies by $\sim 5-20\%$ for $n=0$ and 1-5\%
for $n>0$ compared with the stratified case. The crust increases the
frequency because it increases the average shear modulus of the
star. The effect is small because the shear modulus of the crust
exceeds $\ev{B_t^2}/4\pi$ in only a small fraction of the stellar
volume. The effects of crust rigidity are therefore small compared to
the rigidity introduced by the tangled field throughout the star, and
the $l=1$ modes are unchanged to two significant figures. A simple
analysis is given in Appendix
\ref{crust} that further quantifies this effect. By contrast, the 
rigidity of the crust is essential an essential ingredient if the
field is not significantly tangled, as considered in previous work
(\eg,
\citealt{vl11,ck11,gabler_etal11,gabler_etal12,vl12,gabler_etal13,gabler_etal_sf13,gabler_etal14}).

\begin{table}
\begin{tabular}{l|lllllll}
\hline
$l$ & $n=0$ & $n=1$ & $n=2$ & $n=3$ & $n=4$ & $n=5$ & $n=6$ \\
\hline
1 &  0        & -15\%& -15\%& -16\%& -15\% & -16\% & -16\% \\
2 & +48\%& -8.9\%& -12\%& -14\%& -15\% & -15\%& -15\%\\
3 & +41\%& -4.5\%& -10\%& -12\%& -13\%& -14\%& -14\%\\
4 & +35\%& -1.9\%& -8.0\%& -11\%& -12\%& -13\%& -14\%\\
5 & +31\%& -0.47\%& -6.5\%& -9.4\%& -11\%& -12\%& -13\%\\
6 & +27\%& 0.33 \%& -5.6\%& -8.5\%& -10\%& -11\%& -12\%
\end{tabular}
\caption{Percentage frequency change of modes in a stratified star with a crust compared with a constant density star with the same mass and radius.  We use a 1.4 $M_\odot$ star with radius 13.1 km, taking $x_p=0.1$, and magnetic field $\ev{B_t^2}=10^{15}$ G}.
\label{strat_values_pc}
\end{table}

On the balance, stratification has a larger effect than does the
crust.  The combined effects of stratification and a crust is
presented in Table \ref{strat_values_pc}, where the percentage change
in frequencies compared with a constant density star of 13.1 km radius
are given.  For the $n=0$ modes, the combined effects of
stratification and a crust increase frequencies by up to 50\%, because
both effects add.  For $n>1$ the frequencies drop as a result of
stratification, but are enhanced slightly by the crust giving a net
change of up to 16\%.  Changes become more pronounced for increasing
$n$, but less pronounced for increasing $l$.

We conclude that for star in which the tangled field accounts for much
of the magnetic energy density ($b_t^2\gap 1$), that both density
stratification and the crust rigidity have a relatively small effect
on the frequencies of torsional normal modes. The constant-density
model is an accurate first approximation at the 10\% level for most
overtones, and better than 50\% for fundamental modes.

\section{Energetics}

\label{energetics}

The only quantitative discussion to date of the energetics of QPO
excitation has been given by \citet{lv11}. We use the results of their
analysis to calculate the mode amplitude at the stellar surface in the
global oscillation interpretation of QPOs, and the total energy in the
low-frequency QPOs.

The natural candidate for the energy source that excites stellar modes is
the flare itself. For giant flares, the energy cannot be released deep
inside the star, as the impedance mismatch between the stellar
interior and the magnetosphere is so great that the energy in
Alfv\'en waves would take
seconds or longer to reach the magnetosphere due to multiple
reflections, in conflict with observed rise times of $\lap 10$ ms
\citep{link14_flares}. 
This constraint favors an external origin for
flares, as suggested by
\citet{lyutikov03,lyutikov06}, \citet{komissarov_etal07}, and \citet{gh10}. 
Forcing of the magnetosphere through episodic release
of {\em internal} magnetic energy could produce a magnetospheric
explosion when an MHD instability is triggered. 

\citet{lv11} considered the excitation problem under the assumption
that the magnetosphere undergoes a sudden change (over several
relativistic Alfv\'en wave crossing times in the magnetosphere, $\sim
10^{-4}$ s) from one equilibrium configuration to another, exciting
torsional oscillations in the star. Suppose the shear stress at the
surface changes by $fB^2/4\pi$, where $B$ is the characteristic
field strength in the inner magnetosphere, and $f\le 1$.  The average
displacement $\bar{u}_i$ at the stellar surface, assuming excitation
of {\em global} mode $i$ with frequency $\omega_i$, is of order
(\citealt{lv11}; eq. 11)
\begin{equation}
\frac{\bar{u}_i}{R}\sim \frac{1}{4\pi}\frac{ fB^2 R}{M\omega_i^2} =
\frac{3}{2\pi}\frac{fE_{\rm mag}}{MR^2\omega_i^2}=
10^{-3}\, \left(\frac{fE_{\rm mag}}{2\times 10^{47}\mbox{ erg}}\right)
\left(\frac{R}{10\mbox{ km}}\right)
\left(\frac{M}{1.4M_\odot}\right)^{-1}
\left(\frac{\nu_i}{\mbox{30 Hz}}\right)^{-2},
\label{u}
\end{equation}
where $E_{\rm mag}\equiv (4\pi R^3/3)(B^2/8\pi)\sim 2\times 10^{47}
B_{15}^2$ erg is the characteristic magnetic energy available to be
released; $fE_{\rm mag}$ is the total energy released, and should be
comparable to the flare energy.  The scaling as $\omega_i^{-2}$
follows from the Fourier transform of a step function.

The energy in a mode is of order (\citealt{lv11}; eq. 12) 
\begin{equation}
E_i\sim \frac{f^2B^4R^4}{32\pi^2 M \omega_i^2}.
\end{equation}
Most of the energy is in low-frequency modes. We have found
a fairly dense spectrum with a characteristic mode spacing of
$\Delta\nu\simeq 10$ Hz. The sum of the total energy in all modes can
be approximated with an integral
\begin{equation}
E\simeq \int_{\nu_0}^\infty \, d\nu\, \frac{E(\nu)}{\Delta\nu}\sim
10^{44}\, f^2\left(\frac{B}{10^{15}\mbox{ G}}\right)^4
\left(\frac{R}{10\mbox{ km}}\right)^4
\left(\frac{M}{1.4M_\odot}\right)^{-1}
\left(\frac{\nu_0}{20\mbox{ Hz}}\right)^{-1}
\left(\frac{\Delta\nu}{10\mbox{ Hz}}\right)^{-1}
\mbox{ erg}, 
\end{equation}
where $\nu_0$ is the lowest-frequency normal mode in the spectrum. We
see that the conversion of magnetospheric energy to mechanical motion
in the star an inefficient process; about 0.1\% of the flare energy
goes into normal modes. The total energy in the modes in our model is
well within the energy budget of a giant flare.

For a small flare, with $fE_{\rm
mag}=10^{39}$ erg, the amplitude from eq. \ref{u} is smaller by $\sim
10^8$ than if $E_{\rm mag}$ is liberated.  These estimates hold not
just for our model, but generally for any global-oscillation model of
QPOs.

\section{Discussion and Conclusions}

\label{conclusions}

We propose a new physical scenario for magnetar oscillations wherein
the Alfv\'en continuum supported by the organized field is broken by
nearly isotropic stresses arising from components of the field that
form a complex tangle over scales much smaller than the stellar
radius. In a simple model of constant density, an organized field of
strength $\sim 10^{15}$ G, and a tangled field of comparable strength,
a fundamental frequency appears near 20 Hz for $l=1$ and $l=2$, with a
with a mode spacing of $\sim 10$ Hz (we use $l$ from the solutions of
the isotropic problem to label the modes of the anisotropic
problem). The model is consistent with QPOs observed in SGRs 1806 and
1900 below about 160 Hz. Magnetic stresses dominate material stresses
almost everywhere in the crust, so crust elasticity is unimportant
when a strong, tangled field is present. We find that the combined
effects of realistic stellar structure and crust stress give a modest
change in the eigenfrequencies of $\sim 10$\% for most modes (but up
to 50\% for low-$l$ fundamentals) so
our simple model with constant density is a good first approximation
to studying tangled fields that elucidates the essential physics. The
strength of the organized field, which should be approximately
dipolar, is determined observationally. Once stellar mass and radius
are fixed, the model has only one important free parameter: the ratio
of the energy densities in tangled and organized fields.

Our results are insensitive to whether or
not the protons are superconducting. The upper critical field for
type II superconductivity is
$\lap 10^{16}$ G, comparable to average fields we find to be
consistent with the QPOs observed in SGRs 1806 and 1900; the
magnetic stress is nearly the same if the protons are superconducting,
though we have assumed for definiteness that they are normal. Also, if
the protons are superconducting, the neutrons are only slightly
entrained by the protons throughout most of the core, and the
dynamical mass density is nearly equal to the proton mass density in
this case as well.

While our constant-density model is mostly illustrative, it is
nevertheless sufficiently accurate to study torsional oscillations
semi-quantitatively within our approximations; the model is able
to account for {\em every} QPO observed in SGRs 1806 and 1900 to
within 3 Hz. No other published model has given such quantitative
agreement, though we acknowledge that the success of the model is
due in large part to the rather dense spectrum we predict with a
frequency spacing of about 10 Hz. We find the data are best
explained if there is approximate equipartition of the energy in the
smooth and tangled fields. Such modes have high enough wavenumbers to
probe the field structure over scales below $0.2R$, at which point the
averaging procedure we introduced to treat the tangled field might
become inappropriate.

Why have most of the predicted frequencies not been observed?  One
possibility is that the modes have been excited, but are not
visible. Given that we do not understand how surface oscillations
affect magnetospheric emission, it could be that some modes are not
seen because they do not produce sufficiently large changes in the
x-ray emissivity to be visible from our viewing angle. Another
possibility is that the instability that drives the flare creates
highly-preferential excitation of axial modes. We note that
preferential excitation of low-frequency free oscillations occurs in
the Earth \citep{rr04}, the so-called ``hum''. The question of how
preferential excitation of stellar modes might occur is an interesting
question.

While our focus has been on SGRs 1806 and 1900, which have dipole
fields of $\sim 10^{15}$ G, QPOs have been reported for
SGR J1550-5418 \citep{huppenkothen_etal14b} in association with burst
storms. SGR J1550-5418 has an inferred
dipole field of $3.2\times 10^{14}$ G \citep{camilo_etal07}. In this
object, as in SGR 1900, the observed QPOs can be accounted for if most
of the magnetic energy is in the tangled field, so that the average
magnetic field is $\sim 10^{15}$ G.

A crucial ingredient in the interpretation of QPOs as stellar
oscillations is to understand how crust movement can produce the large
observed modulations of the x-ray emission by
10-20\%. \citet{timokhin_etal08} propose that twisting of the crust,
associated with a stellar mode, modulates the charge density in the
magnetosphere, creating variations in the optical depth for resonant
Compton scattering of the hard x-rays that accompany the flare. In
that model, the shear amplitude at the stellar surface must be as
large as 1\% of the stellar radius. \citet{dw12} accounted for
geometrical effects of the beamed emission, and find that the
amplitude of the QPO emission is increased by a factor of typically
several over the estimate of \citet{timokhin_etal08} for a given
amplitude of the surface displacement.  Based on the analysis of
\cite{lv11} of mode excitation by a sudden change in magnetospheric 
equilibrium, we find that the predicted, low-frequency QPO spectrum
could contain up to $\sim 10^{44}$ erg in mechanical energy. This
number is well within the energy budget of giant flares but the
surface amplitude is only $\sim 10^{-3} R\, (30\mbox{ Hz}/\nu)^2$, 
smaller than required by
\citet{timokhin_etal08}, especially for the QPOs near 160 Hz. 
This small amplitude could spell trouble for the
global-oscillation interpretation of QPOs, unless a more efficient
mechanism than that of
\citet{timokhin_etal08} can be identified. Explaining QPOs in small bursts
($\sim 10^{39}$ erg) appears particularly difficult with a global
oscillation model. The mode amplitude scales as the flare energy
(eq. \ref{u}), giving a surface amplitude that is $\sim 10^8$ times smaller
than for a giant flare. The 626 Hz QPO
reported in SGR 1806 also seems problematic to excite in a giant
flare, with a surface amplitude of $\sim 10^{-6}R$ in the most optimistic
case. These numbers show that energetics should be given serious
consideration in QPO models, and that other excitation mechanisms than
that of \citet{lv11} should be considered.

The simple model presented in this paper can be improved by including
a crust and introducing realistic stellar structure. We have shown
that the effects typically $\sim 10$ for most modes for a star in
which the magnetic energy density is dominated by that in the tangled
field; the effects of stellar structure and the crust will become more
important as the effective shear modulus from the tangled field is
reduced, and we have not quantified these effects. Other modes than
axial modes, such as polar modes, should be considered as well.

An explanation for why the observed frequencies are quasi-periodic is
well beyond the scope of our model. Quasi-periodicity could arise from
magnetospheric effects, dynamics in the star not included here, or
both. 

\section*{Acknowledgments}

We thank M. Gabler, D. Huppenkothen, Y. Levin, and A. Watts for very
helpful discussions, and Y. Levin and A. Watts for comments on the
manuscript.  This work was supported by NSF Award AST-1211391 and NASA
Award NNX12AF88G.

\appendix

\section{Variable Separation}

\label{sov}

Eqs. (\ref{eomt}) and (\ref{bct}) can be separated and solved by
transforming to an oblate spheroidal coordinate system $(u,v)$ defined
by
\begin{equation}
s=R
\sqrt{\frac{\left(1+b_t^2 u^2\right)\left(1-v^2\right)}
{1+b_t^2}}
\end{equation}
\begin{equation}
z=R u v
\end{equation}
Curves of constant $u$ are ellipses, and curves of constant $v$ are
hyperbolae. For $u=1$, the coordinate gives a sphere of radius
$R$. In the limit $b_t^2\rightarrow \infty$, spherical 
coordinates are recovered with $u=r/R$ and
$v=\cos\theta$. 

In these coordinates, eq. (\ref{eomt}) becomes
\begin{eqnarray}
\left(1+b_t^2 u^2\right)\frac{\partial^2 u_\phi}{\partial u^2}+ && 2 b_t^2 u \frac{\partial u_\phi}{\partial u} +\frac{b_t^2 u_\phi}{1+b_t^2 u^2}  
+b_t^2\left(1-v^2\right)\frac{\partial^2 u_\phi}{\partial
v^2}\nonumber \\
&& -2 b_t^2 v \frac{\partial u_\phi}{\partial v} -\frac{b_t^2u_\phi}{1-v^2}  \nonumber 
+\bar{\omega}^2\left(b_t^2 u^2+ v^2\right)u_\phi=0
\label{eomtuv}
\end{eqnarray}
where
\begin{equation}
\bar\omega\equiv \frac{R \omega}{c_t}\sqrt{\frac{b_t^2}{1+b_t^2}}, 
\end{equation}
and $c_t^2\equiv\ev{B_t^2}/(4\pi\rho_d)$.

The boundary condition eq. (\ref{bct}) at $u=1$ becomes
\begin{equation}
(1+b_t^2)\frac{\partial u_\phi}{\partial u}-b_t^2 u_\phi=0,
\label{bcuv}
\end{equation}
while at $z=0$ we require
\begin{equation}
  u_\phi=0\,, \label{bczodd}
\end{equation}
for odd parity modes or
\begin{equation}
  \left(1+b_t^2 u^2\right) v \frac{\partial u_\phi}{\partial u}+b_t^2\left(1-v^2\right) u \frac{\partial u_\phi}{\partial v}=0\,, \label{bczeven}
\end{equation}
for even parity modes.

We seek a separable solution of the form
\begin{equation}
u_\phi(u,v)=U(u)V(v).
\end{equation}
Eq. (\ref{eomtuv}) becomes
\begin{equation}
   \left(1+b_t^2 u^2\right)\frac{d^2 U}{d u^2}+2 b_t^2 u \frac{d U}{d u} +\frac{b_t^2 U}{1+b_t^2 u^2} -\kappa b_t^2 U +\bar{\omega}^2 b_t^2 u^2 U=0
\label{sepu}
\end{equation}
\begin{equation}
b_t^2 \left(1-v^2\right)\frac{d^2 V}{d v^2}-2 b_t^2 v \frac{d V}{d v} -\frac{b_t^2 V}{1-v^2} +\kappa b_t^2 V + \bar{\omega}^2 v^2 V=0,
\label{sepv}
\end{equation}
where $\kappa$ is the separation constant. The boundary condition
eq. (\ref{bcuv}) becomes
\begin{equation}
(1+b_t^2)\frac{dU}{du}-b_t^2U=0,
\label{bcu}
\end{equation}
at $u=1$. 
The boundary conditions (\ref{bczodd}) and (\ref{bczeven}) are satisfied if we impose
\begin{equation}
  U(0)=0 \hspace{5mm} and \hspace{5mm} V(0)=0\,,
\end{equation}
for odd parity modes or
\begin{equation}
  U'(0)=0 \hspace{5mm} and \hspace{5mm} V'(0)=0\,,
\end{equation}
for even parity modes.

In the limit $b_t^2\rightarrow
\infty$, eq. (\ref{sepu}) reduces to the spherical Bessel equation
with solution $U(u)=j(\bar{\omega} u)$, while (\ref{sepv}) reduces to
an associated Legendre equation 
with solution
$V(v)=\sqrt{1-v^2}\partial P^0_{l}/\partial v=-P^1_{l}(v)$ and
$\kappa=l(l+1)$, which recovers the solutions of the isotropic model of
\S \ref{isotropic_model}.

We solve eqs. (\ref{sepu}), (\ref{sepv}), (\ref{bcu}) numerically for
the normal mode frequencies.

For arbitrary  $b_t^2$ the modes are identified by the number of
nodes; for the $n^{th}$ mode the function $U(u)$ has $n$ nodes on the
domain $0\leq u \leq 1$, while for the $l^{th}$ mode the function
$V(v)$ has $l-1$ nodes on $-1\leq v \leq 1$.

\appendix

\setcounter{section}{1}

\section{Effects of the Crust}

\label{crust}

The comparisons to observed QPOs in \S \ref{anisotropic_model} suggest
$b_t^2\sim 1$, so that $\ev{B_t^2}/4\pi\sim B_o^2/4\pi\simeq 8\times
10^{28}$ \ergcc. The shear modulus crust exceeds this value only in
the densest regions of the crust. The rigidity of the crust will
increase the normal-mode frequencies somewhat with respect to what we
have found by neglecting the crust. Here we show that crust rigidity
has small or negligible effects on our results.

To estimate the effect, we consider a two-component isotropic model
with only a tangled field. The core liquid, which we take to be
homogeneous, has an effective shear modulus $\mu_{\rm
core}\equiv\ev{B_t^2}/4\pi$ and a shear-wave speed $c_t$. The crust,
which we also take to be homogeneous, has shear modulus $\mu_{\rm
crust}$ and a shear-wave speed $c_\mu$. The crust has an inner radius
$R_c$ and outer radius $R$. 

In the core, the solution to the mode
problem is $u_{\rm core}=j_l(kr)$; the mode frequency is
$\omega=c_tk$. The solution in the crust is $u_{\rm
crust}=aj_l(k^\prime r)+bn_l(k^\prime r)$, where $n_l$ are spherical
Neumann functions, $a$ and $b$ are constants, and $\omega=c_\mu
k^\prime$. The boundary conditions are continuity in value and
traction at $r=R_c$, and vanishing traction at $r=R$:
\begin{equation}
u_{\rm core}(R_c)=u_{\rm crust}(R_c),
\end{equation}
\begin{equation}
\mu_{\rm core}\left[\frac{du_{\rm core}}{dr}-\frac{u_{\rm core}}{r}
\right]_{r=R_c}=
\mu_{\rm crust}\left[\frac{du_{\rm crust}}{dr}-\frac{u_{\rm crust}}{r}
\right]_{r=R_c},
\end{equation}
\begin{equation}
\mu_{\rm crust}\left[\frac{du_{\rm crust}}{dr}-\frac{u_{\rm crust}}{r}
\right]_{r=R}=0.
\end{equation}

The shear modulus in the crust, ignoring magnetic effects, is
\citep{strohmayer_etal91} 
\begin{equation}
\mu=\frac{0.1194}{1+0.595(173/\Gamma)^2}\frac{n_i(Ze)^2}{a}, 
\end{equation}
where $n_i$ is the number density of ions of charge $Ze$, $a$ is the
Wigner-Seitz cell radius given by $n_i4\pi a^3/3=1$, and $\Gamma\equiv
(Ze)^2/(akT)$ where $k$ is Boltzmann's constant. Typically in the
crust, $\Gamma>>173$ and the second term in the denominator is
negligible. For the composition of the inner crust, we use the results of
\citet{dh01}, conveniently expressed analytically
by \citet{hp04}. We solve
for crust structure using the Newtonian equation for hydrostatic
equilibrium, for a stellar radius of 10 km and a stellar mass of 1.4
$M_\odot$. In the evaluation of the shear-wave speed in the crust, we
include the effects of nuclear entrainment
\citep{chamel05,chamel12}. Further details are given in
\citet{link14_flares}. 

We find that $\mu_{\rm crust}$ drops from $2\times 10^{30}$ \ergcc\ at
the base of the crust to $8\times 10^{28}$ \ergcc\ over 370 m. Because
we have assumed a homogeneous crust, we take the geometric mean of
$\mu_{\rm crust}$ over this range, which is $4\times 10^{29}$
\ergcc, and set the crust thickness to 370 m. (The matter in the
remainder of the crust contributes less to the rigidity than does the
tangled field). The shear speed in the crust varies by a factor of about two
in this region, with a geometric mean of $5\times 10^{-3}c$. With
these values, we solve the boundary conditions numerically for the
corrected eigenfrequencies. Crust rigidity increases the mode
frequency by $\lap 12$\% for each fundamental of a given $l$, and
$\lap 3$\% for harmonics.

The treatment by \cite{dh01} of the inner crust gives
somewhat higher values of the shear modulus at the base of the crust
than do other studies. The equation of state
of \cite{akmal_etal98}, for example, gives a shear speed at the base of the
crust that is about 0.6 the shear speed 
of \citet{dh01}, and a corresponding shear modulus that is smaller by
a factor of about 2.8. Reducing the crust shear modulus by a factor of
two, and repeating the above analysis, shows that crust rigidity
increases the fundamental frequency for 
each $l$ by less than 10\%.


\label{lastpage}

\end{document}